\documentclass[twocolumn,a4paper]{article}

\usepackage{cite}
\usepackage{graphics}
\usepackage{amsmath}
\usepackage{amsfonts}
\usepackage{amssymb}
\usepackage{epsfig}

\begin{document}

\title{Depairing currents in the superconductor/ferromagnet proximity system
Nb/Fe}
\author{J. M. E. Geers, M. B. S. Hesselberth, J. Aarts \thanks{corresponding author; email
aarts@phys.leidenuniv.nl} \ \\ \emph{Kamerlingh~Onnes~Laboratory,
Leiden Institute of Physics,}  \\ \emph{Leiden~University,
P.O.~Box~9504, NL-2300~RA~Leiden, The~Netherlands} \\ A. A.
Golubov \\ \emph{University of Twente, P.O.~Box 217, NL-7500 AE
Enschede, the Netherlands} }
\date{\today}

\maketitle

\begin{abstract}
We have investigated the behaviour of the depairing current
$J_{dp}$ in ferromagnet/superconductor/ferromagnet (F/S/F)
trilayers as function of the thickness $d_s$ of the
superconducting layers. Theoretically, $J_{dp}$ depends on the
superconducting order parameter or the pair density function,
which is not homogeneous across the film due to the proximity
effect. We use a proximity effect model with two parameters
(proximity strength and interface transparency), which can also
describe the dependence of the superconducting transition
temperature $T_c$ on $d_s$. We compare the computations with the
experimentally determined zero-field critical current $J_{c0}$ of
small strips (typically 5~$\mu$m wide) of Fe/Nb/Fe trilayers with
varying thickness $d_{Nb}$ of the Nb layer. Near $T_c$ the
temperature dependence $J_{c0}(T)$ is in good agreement with the
expected behaviour, which allows extrapolation to $T$ = 0. Both
the absolute values of $J_{c0}(0)$ and the dependence on $d_{Nb}$
agree with the expectations for the depairing current. We conclude
that $J_{dp}$ is correctly determined, notwithstanding the fact
that the strip width is larger than both the superconducting
penetration depth and the superconducting coherence length, and
that $J_{dp}(d_s)$ is correctly described by the model.
\end{abstract}

\renewcommand{\thesection}{\Roman{section}}
\section{Introduction}
A still relatively little explored area of research in
non-equilibrium superconductivity concerns phenomena involving
spin-polarized quasiparticles. Pioneering work on spin-polarized
tunneling in conventional s-wave superconductors was performed by
Meservey and Tedrow \cite{mes94}, who studied different
ferromagnets (F) in F/Al$_2$O$_3$Al tunnel junctions and found
that the tunnel current can show varying degrees of spin
polarization. More recently, experiments were performed by
different groups in order to establish whether superconductivity
can be suppressed by injecting spin-polarized quasiparticles
\cite{vasko97,dong97,yeh99}. In these cases the combinations
existed of a d-wave high-T$_c$ superconductor (XBa$_2$Cu$_3$O$_7$,
with X = Y, Dy) and a fully spin-polarized ferromagnetic manganite
(A$_{0.67}$B$_{0.33}$MnO$_3$ with A = La, Nd and B = Ca, Sr),
either with or without a barrier of a different oxide; measured
was the change in the zero-field critical current density $J_{c0}$
of the superconducting films upon applying a current bias through
the ferromagnet. The results are not fully conclusive, and
certainly not quantitative. Although generally a suppression of
$J_{c0}$ was observed, heating effects could not always be fully
ruled out since the manganites are highly resistive metals (see
the discussion in ref. \cite{yeh99}), and the geometry did not
always allow to determine the area of the current injection, and
therefore the injected current density. Moreover, since $J_{c0}$
in high T$_c$ superconductors generally is not the depairing
current $J_{dp}$ but involves flux motion, $J_{c0}$ is not a
direct measure for the amount of depression of the superconducting
order parameter. Similar experiments have to our knowledge not
been performed with combinations of conventional metals, although
that would have some clear advantages. The interpretation of
results would not be complicated by e.g. inhomogeneous currents in
the ferromagnet or anisotropic gaps in the superconductor;
lithographic techniques could be brought to bear in order to have
well defined superconducting bridges and injection contacts; and
it should be possible to identify the effects of the
spin-polarized quasiparticles on $J_{dp}$. \\

\noindent Still, two points deserve special interest. The first is
that, in planning such an F/I/S experiment, there is the potential
problem of insufficient knowledge of the tunneling process. This
was already apparent in the work of Meservey and Tedrow cited
above \cite{mes94} since the experiments always showed a positive
sign for the spin polarization, even in the cases of e.g. Co and
Ni where a negative sign was expected. Recently, this was
explained by demonstrating that the choice of barrier material can
strongly influence and even reverse the spin polarization of the
tunneling current \cite{teresa99}, with obvious consequences for
the interpretation of the injection experiments. It may be
advantageous to also contemplate an (F or N)/I/F/S configuration;
in this case the barrier is only used to increase the energy of
the electrons coming from an N or F contact, while the
polarization now takes place in a thin F layer between barrier and
superconductor. The disadvantage here is that the F layer in
connection with the superconductor will suppress the order
parameter and therefore $J_{dp}$ in the S layer. Still, since the
proximity effect (which gives rise to the depression) for S/F
systems is understood reasonably well \cite{aarts97}, the effect
on $J_{dp}$ may also be quantifiable. The second point for
consideration is that even in the case of conventional
superconductors the determination of $J_{dp}$ need not be
straightforward. The difficulty lies in the fact that the
superconducting bridge must have a width of no more than both the
superconducting penetration depth $\lambda$ and the
superconducting coherence length $\xi$. The first is needed to
avoid current pile-up near the edges (as a consequence of
screening of the self-field), the second is required in order to
avoid vortex nucleation and flow, which gives rise to dissipation
before $J_{dp}$ is reached. These conditions can be met e.g. for
Al, which has a Bardeen-Cooper-Schrieffer (BCS) coherence length
$\xi_0$ of about 1.5~$\mu m$, while $\lambda$ can also be made of
the order of 1~$\mu m$ by making the film thin enough. For
Al-bridges of less than 1 $\mu m$ wide it was shown by Romijn {\it
et al.} \cite{romijn82} that the measured $J_{dp}$ agreed very
well with the theoretical calculations by Kuprianov and Lukichev
\cite{kupr80} based on the Eilenberger equations and therefore
valid in the whole temperature regime below $T_c$. For a material
such as Nb, with $\xi_0$ and $\lambda$ of the order of 50~nm, such
agreement need not be expected. \\

\noindent In this paper we show that, at least close to $T_c$, the
values of the zero-field critical current $J_{c0}$ measured on
bridge-structured Nb samples are essentially the values expected
for the depairing current. Furthermore, we measure the depression
of $J_{c0}$ in trilayers of Fe/Nb/Fe as function of the thickness
$d_{Nb}$ of the Nb layer. We compare the behaviour of
$J_{c0}(d_{Nb})$ with the behaviour of $T_c(d_{Nb})$, and also
with calculations of the proximity effect and the pairbreaking
velocity using a two-parameter formalism based on the Usadel
equations. We find that $J_{c0}(d_{Nb})$ is well described by the
same two parameters which describe the behaviour of $T_c(d_{Nb})$.
The conclusion is that the suppression of the depairing current as
a consequence of the depression of the order parameter in S/F
structures can be well described by proximity effect theory,
making (F,N)/I/F/S injection experiments a distinct possibility.

%%%%%%%%%%%%%%%%%%%%%%%%%%%%%%%%%%%%%%%%%%%%%%%%%%%%%%%%%%%%%%%%%%

\section{Depairing current; theory}
\label{theo} 
\begin{figure}
\epsfig{file=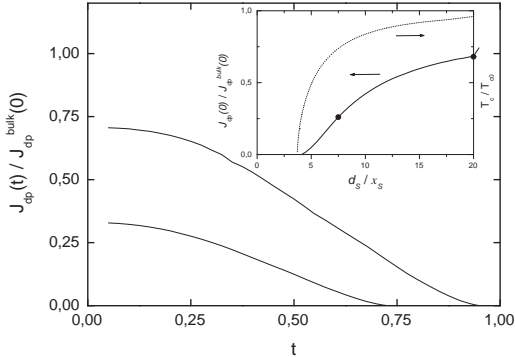,width=\columnwidth} \caption{The temperature
dependence of the normalised depairing current
$J_{dp}(t)/J_{dp}^{\text{bulk}}(0)$ of an F/S/F trilayer for
S-layer thicknesses $d_S / \xi_S$ = 20 (upper), 7.5 (lower).
Parameters typical for Fe/Nb/Fe were used, namely $\gamma = 34.6$
and $\gamma_b = 42$. Insert : thickness dependences of the
normalised depairing current at T = 0, and of $T_c$ of the same
trilayer. The black dots are at $d_S / \xi_S$ = 20, 7.5. }
\label{fig:theory}
\end{figure}

Close to $T_c$, the classical Ginzburg-Landau (GL)
result for the temperature dependence of the depairing current of
a thin film, under the assumption of a homogeneous superconducting
order parameter over the film thickness, is given by :
\begin{equation}
J_{dp}^{GL}(t) \; = \; J_{dp}^{GL}(0)\; (1 - t)^{3/2} \;,
\end{equation}
with $t = T/T_c$. The prefactor $J_{dp}$ is of the order of
$H_c/\lambda$, with $H_c$ the thermodynamic critical field, and
will be given more precisely below. For arbitrary temperatures,
calculations were performed by Kupriyanov and Lukichev, who
essentially solved the Eilenberger equations for a superconductor
carrying a current, with the velocity of the condensate leading to
a phase gradient \cite{kupr80}. Their results recover the GL
behaviour near $T_c$ :
\begin{eqnarray}
J_{dp}^{GL}(t) & = & \frac{16}{9\sqrt{7 \zeta(3)}}
\sqrt{\chi(\rho_d)} \; [ e N(0) v_F k_B T_c ] \nonumber \\
 & & \times  \; (1 - t)^{3/2}.
\end{eqnarray}
Here, the constants have their usual meaning, $N(0)$ is the
density of states at the Fermi level per spin direction, and
$\chi(\rho_d)$ is the G'orkov function controlled by the 'dirt
parameter' $\rho_d = (\hbar v_F) / ( 2 \pi k_B T_c \ell_e )$, with
$\ell_e$ the electronic mean free path. In the dirty limit, (
$\rho_d \rightarrow \infty$) $\chi(\rho_d) \rightarrow 1.33 \ell_e
/ \xi_0$, this becomes :
\begin{equation} \label{eq:jdpgl}
J_{dp}^{GL}(0) \; = \; 1.26 \; [ e N(0) v_F \Delta(0) ] \;
\sqrt{\frac{\ell_e}{\xi_0}}
\end{equation}
which is equivalent to the expression given by Romijn et al.
\cite{romijn82} :
\begin{equation} \label{romijn}
J_{dp}^{GL}(0)  = \frac{16 \pi^2 \sqrt{2 \pi}}{63 \zeta(3)} [ e
N(0) v_F k_B T_c ] \sqrt{\frac{k_B T_c \ell_e}{\hbar v_F}} ,
\end{equation}
which can also be written in terms of experimental parameters as
\begin{equation}
J_{dp}^{GL}(0) = 7.84 \left[ \frac{(k_B T_c)^3}{e^2 \hbar v_F
(\rho \ell)} \frac{1}{\rho}\right]^{1/2}. \label{jdp2}
\end{equation}
This way of writing also emphasizes the proportionality
$J_{dp}^{GL}(0) \propto \sqrt{1/\rho}$, since the product $\rho
\ell$ is a materials constant. At low temperatures the value of
$J_{dp}$ saturates, reaching a zero-temperature value of :
\begin{eqnarray} \label{eq:delta32}
J_{dp}(0)& = & 1.491 e N(0) \sqrt{\frac{D}{\hbar}} \Delta^{3/2}(0)
\nonumber \\
         & = & 0.486 \; [ e N(0)v_F \Delta(0) ] \;
\sqrt{\frac{\ell_e}{\xi_0} } \;,
\end{eqnarray}
with $D = 1/3v_F \ell_e$ the diffusion constant. Comparison with
eq.~\ref{eq:jdpgl} shows that the ratio between the saturation
value and the GL-extrapolated value equals $J_{dp}(0) /
J_{dp}^{GL}(0)$ = 0.385.
\\
In the case of F/S (or N/S) multilayers, the superconducting order
parameter is depressed near the interfaces, and this has to be
taken into account in calculating $J_{dp}$. For this we use the
proximity effect model, based on the Usadel equations (dirty limit
conditions), which was also used for calculating the depression of
$T_c$ with decreasing thickness of the superconductor
\cite{aarts97}. Details will be given in Appendix~A but here we
briefly introduce the main parameters of the theory. In principle,
the shape of the order parameter on both sides of the interface
depends on the bulk transition temperature $T_{c0}$, on the
coherence lengths $\xi_{S,F}$, on the normal state resistivities
$\rho_{S,F}$, and on the transparency $T$ of the interface. From
the boundary conditions for the order parameter (see
eqs.~\ref{eq:bound}) it follows that, apart from $T_{c0}$, only
two independent parameters are needed, the proximity strength
parameter $\gamma$ and the transparency parameter $\gamma_b$. The
value of $\gamma$ = $(\xi_S \rho_S)/(\xi_F \rho_F)$ can be fully
determined from experiment; the only free parameter is $\gamma_b$
($0 \leq \gamma_b \leq \infty$), which is approximately connected
to the transparency $T$ (with $0 \leq T \leq 1$) by
\begin{equation}
T \; = \; \frac{1}{1+\gamma_b}
\end{equation}
As was shown in ref~\cite{aarts97}, in F/S systems, $T$ can be
quite low for a high magnetic moment in the F-layer, which is
presumably due to the suppression of Andreev reflections by the
exchange splitting of the spin subbands. Fig.~\ref{fig:theory}
gives the results of some typical calculations, performed for the
system Fe/Nb/Fe with the appropriate proximity effect parameters
$\gamma =$ 34.6 and $\gamma_b = $ 42 (see section~\ref{results}).
Shown is $J_{dp}(t)$ for two different thicknesses ($d_S / \xi_S =
20, 7.5$), normalised on the bulk value $J_{dp}^{bulk}(0)$ as
given by eq.~\ref{eq:delta32}. Note that this involves a factor
$(T_c / T_c^{bulk})^{3/2}$.
The thickness dependence of $T_c$ and the normalised depairing
current at T = 0 (see the insert of Fig.~\ref{fig:theory}) are
quite different, with a much stronger depression of the depairing
current at relatively high thickness of the superconductor. This
can be qualitatively understood by noting that $T_c$ is a measure
for the maximum value of the superconducting order parameter in
the layer, while the depairing current comes from an average over
the layer thickness, which also involves lower values of the order
parameter.

\section{Experimental} \label{expt}
Samples were grown on Si(100) substrates, by DC sputtering in a
system with a base pressure of 10$^{-9}$~mbar in an Ar-pressure of
$6 \times 10^{-3}$ mbar. Sputtering rates were of the order
0.1~nm/s for Nb and 0.03~nm/s for Fe. One series of samples
consisted of trilayers Nb/Fe/Nb with Nb thickness $d_{Nb}$~= 5~nm
and the Fe thickness $d_{Fe}$ varying between 2~nm and 25~nm.
These were used to determine the magnetization $M_{Fe}$ of the Fe
layers in the presence of Fe/Nb interfaces with a commercial
SQUID-based magnetometer. The behavior of $M_{Fe}$ versus $d_{Fe}$
could be well described with a straight line, yielding a magnetic
moment per Fe atom of 2.36~$\mu_B$ ($\mu_B$ being the Bohr
magneton), slightly above the bulk value of 2.2~$\mu_B$ and a
magnetically dead layer per interface of 0.1~nm. This value is
somewhat lower than reported for MBE-grown samples
\cite{muhge98,verbanck98}. Two other series of samples consisted
of trilayers of Fe/Nb/Fe with $d_{Fe}$~= 5~nm and varying
$d_{Nb}$. One set was structured by Ar-ion etching into strips
with a width $w$ = 100 $\mu m$, the other into strips with a width
$w$~= 6~$\mu m$, or sometimes 10~$\mu m$ or 20~$\mu m$. In both
cases the length between the voltage contacts was 1~mm. The first
set (deposited in two different runs) was used for measuring
$T_c(d_{Nb})$, the second set for both $T_c(d_{Nb})$ and
$J_c(d_{Nb})$. In all cases, the typical width of the resistive
transitions to the superconducting state was 50~mK.

Also measured were single films of Fe and Nb with different strip
widths in order to establish values for the specific resistivity
$\rho_{Fe,Nb}$ (at 10~K), for $T_c$ and for the upper critical
field $B_{c2}(T)$. On average, we find $\rho_{Fe} \approx$
7.5~$\mu \Omega cm$, $\rho_{Nb} \approx$ 3.7~$\mu \Omega cm$,
$T_c$~= 9~K and $S$~= $-dB_{c2} \, / \, dT$~= 0.24 T/K, yielding
$\xi_{GL}(0) =$ $\sqrt{\Phi_0 / (2 \pi S T_c)}$~= 12.2~nm. This
corresponds to $\xi_S$~= 7.8~nm. No special precautions were taken
to shield residual magnetic fields. The zero-field critical
current $I_{c}$ was determined at different temperatures $T$ by
measuring current ($I$)- voltage ($V$) characteristics. For this,
a DC current was switched on for the time of the order of 1~s and
the voltage recorded, to prevent heating via the contacts. All
samples showed a clear transition from the superconducting to the
normal state, with a large and almost instantaneous increase in
voltage at $I_{c}$. Upon detecting this rise, the current was also
turned off since the sample then started to heat immediately. Most
samples also showed a small rise in voltage prior to the major
transition, probably due to vortex motion. We shall come back to
this point in the discussion. Important for the theoretically
expected value of $J_{dp}(0)$ is the value of the resistivity of
the superconducting layer (see Eq.~\ref{romijn}). This value,
$\rho_{Nb}$, was extracted from the normal state resistance $R_n$
at 10~K of the patterned samples by assuming that the Nb layer and
the 10~nm thick Fe layer ($\rho_{Fe}$ = 7.5~$\mu \Omega cm$)
contribute as parallel resistors.
\begin{table}
\begin{center}
\begin{tabular}{@{}c@{\hspace{1mm}}c@{\hspace{1mm}}c@{\hspace{1mm}}c
@{\hspace{1mm}}c@{\hspace{0.1mm}}l@{}} type &$d_{Nb}$ &$w$ &$T_c$
&$\rho_{Nb}$ &$J_{c0}^{GL}(0)$   \\
   &[nm] &[$\mu$m]  &[K]  &[$\mu\Omega$cm] &
 \small [$10^{11}$ A/m$^2$] \normalsize \\
  \hline
F/S/F   &36   &6    &3.63  &5.97  &0.522  \\
F/S/F   &40   &6    &4.36  &6.51  &1.55   \\
F/S/F   &42   &10   &5.07  &10.4  &1.58    \\
F/S/F   &53   &10   &5.62  &8.08  &2.64   \\
F/S/F   &60   &6    &6.63  &5.03  &3.46   \\
F/S/F   &75   &6    &7.34  &4.95  &6.14    \\
F/S/F   &100  &6    &8.05  &4.58  &6.86   \\
F/S/F   &150  &6    &8.61  &3.94  &11.2   \\
Nb      &53   &20   &9.00  &7.24  &15.1   \\ \hline
\\
\multicolumn{5}{l}{$S(Nb)=-\mu_0 \: \mbox{d}H_{c2}/\mbox{d}T
=0.24$~T/K}\\ \multicolumn{2}{l}{$\xi_{\text{Fe}}=0.14$~nm}
&\multicolumn{3}{l}{$\rho_{\text{Fe}}=7.52$~$\mu\Omega$cm}\\
\multicolumn{2}{l}{$\gamma\approx34.6$}
&\multicolumn{3}{l}{$\gamma_b=42$}
\end{tabular}
\end{center}
%\vspace*{0.5cm}
\caption{Parameters of the Fe/Nb/Fe samples and the single Nb film
used for the critical current measurements. Given are the
thickness of the Nb layer $d_{Nb}$, the strip width $w$, the
critical temperature $T_c$, the calculated specific resistance of
the Nb layer $\rho_{Nb}$ and the Ginzburg-Landau extrapolated
critical current at zero temperature $J_{c0}^{GL}(0)$.  }
\label{param}
\end{table}
The resulting values for $\rho_{Nb}$ are given in
Table~\ref{param}, together with the strip width $w$ and $T_c$.
The values for the thinner films (around 50~nm) are somewhat
larger than what we usually find for single Nb films, and approach
that value for the thick films.

\section{Results} \label{results}
\begin{figure}
\epsfig{file=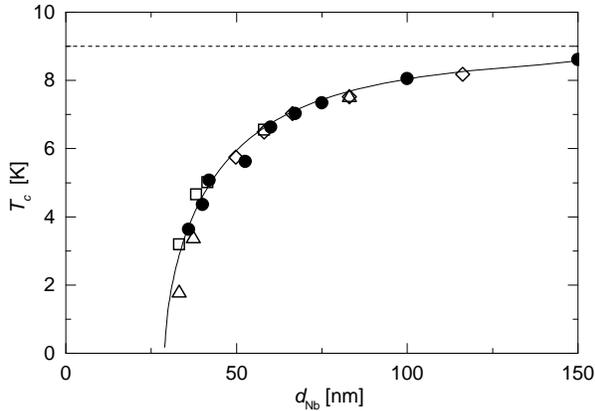,width=\columnwidth} \caption{$T_c$ of the
different sets of Fe/Nb/Fe trilayers. The solid symbols denote the
samples used for the critical current measurements. The line shows
the theoretical dependence $T_c(d_s)$ for the parameter values
$\gamma=34.6$ and $\gamma_b = 42$.} \label{TcdSIc}
\end{figure}

Fig.~\ref{TcdSIc} shows the measured values for $T_c(d_{Nb})$ for
both sample sets, with the the two types of open symbols denoting
the two deposition runs for that set, and the solid symbols
denoting the samples used for measuring $J_{c0}$.
The overall data spread is small, and the data can be well
described by the proximity effect theory for S/F systems we used
for analyzing the behavior of V/(Fe$_x$V$_{1-x})$ in
ref.~\cite{aarts97}, with the two parameters $\gamma$ and
$\gamma_b$ defined above. We use the same value for $\xi_F$ as in
the case of V/Fe, $\xi_{Fe}$ = 0.14~nm and values for $\xi_s$,
$\rho_F$ and $\rho_S$ as given in section~\ref{expt}, yielding
$\gamma$~= 34.6. The best description for $T_c(d_{Nb})$ then is
for $\gamma_b$~=~42, as shown by the drawn line in
Fig.~\ref{TcdSIc}. The critical thickness for the S-layer for
onset of superconductivity $d^S_{cr}$ can be taken either from the
lowest measured value for $T_c$ or from the extrapolated value of
the calculated curve, $d^S_{cr}$~= 29~nm, corresponding to a ratio
$d^S_{cr} / \xi_S$~=~3.7, which is somewhat higher than in the
case of V/Fe where we found 3.2. Apparently, the effect of
ferromagnet on superconductor is slightly stronger in the Nb/Fe
case, but this is not the issue of the current paper.

\begin{figure}
\epsfig{file=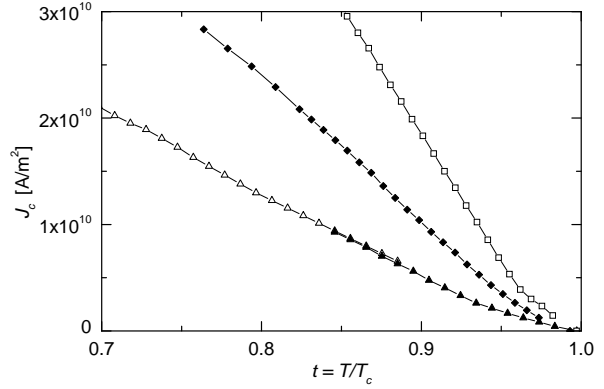,width=\columnwidth} \caption{Experimentally
determined critical current density $J_{c0}$ versus reduced
temperature $t=T/T_c$ for the Fe/Nb/Fe trilayers with
$d_{Nb}=42$~nm (triangles), 60~nm (solid diamonds) and 75~nm (open
squares). The solid and open symbols for $d_{Nb}=42$~nm
corresponds to measurements with non-pumped and pumped He-bath
respectively.} \label{Jcvst}
\end{figure}
\begin{figure}
\epsfig{file=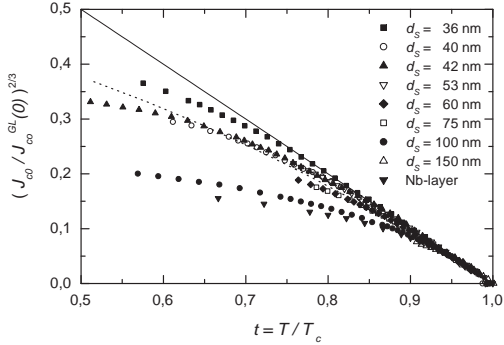,width=\columnwidth}
\caption{$(J_{c0}/J_{c0}^{GL}(0))^{2/3}$ versus $t=T/T_{c}$ for
Fe/Nb/Fe trilayers with different thickness $d_s$ of the Nb-layer,
as indicated. The drawn line indicates the GL-behavior, the dotted
line is the result of the full calculation.} \label{JcJc0vT}
\end{figure}

In Fig.~\ref{Jcvst} $J_{c0} = I_{c0}/(wd)$ is plotted versus
reduced temperature $t=T/T_c$ for $d_{Nb} = 42$, 60 and 75~nm. All
curves show a clear upturn with decreasing temperature in the
region close to $T_c$, above $t \approx 0.9$. Plotting
$J_{c0}^{2/3}(t)$ versus $t$ results in a straight line in this
temperature regime, which can be extrapolated to $t = 0$. The
ensuing values for $J_{c0}^{GL}(0)$ are given in Table~\ref{param}
for all samples, and comprise some of the main experimental
results. They can also be used to normalize the data.
Fig.~\ref{JcJc0vT} shows $[J_{c0}(t)/J_{c0}^{GL}(0)]^{2/3}$ versus
$t$ together with the line $1-t$ (the GL-behavior) and the result
of the full theoretical calculation, which is now independent of
the parameters. All data collapse on the universal curve above $t
= 0.9$. At lower temperatures, the thinnest films ($d_s = $ 36,
40, 42. 53~nm) follow the full calculation quite closely, even
down to $t \approx$ 0.6. The difference between the data of 36~nm
and 40~nm is mainly due to the choice of the normalization value,
and reflects the accuracy of that determination. For thicker films
the first deviation progressively shifts to higher $t$.

\section{Discussion}
\begin{figure}
\epsfig{file=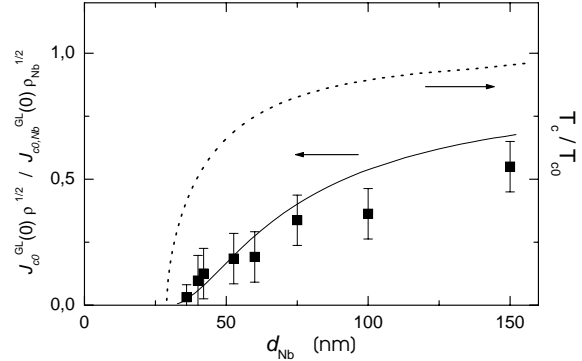,width=\columnwidth} \caption{$J_{c0}^{GL}(0)
\rho^{1/2}$ of the Fe/Nb/Fe trilayers scaled on the value of the
single Nb layer versus superconducting layer thickness $d_{Nb}$.
The result of the model calculations for $\gamma = 34.6$,
$\gamma_b = 42$ is also plotted (solid line) as well as the
dependence of the critical temperature $T_c/T_{c0}$ on $d_{Nb}$
(dashed line) for the same parameters.} \label{grafiek}
\end{figure}

The first point to be discussed is whether the measured values of
$J_{c0}$ agree with the theoretical estimates for $J_{dp}$. The
absolute value of $J_{dp}^{GL}(0)$ can be calculated with
Eq.~\ref{jdp2}. The materials constants for Nb are well documented
\cite{weber91}; we use the values $v_F$~= $5.6 \times 10^5$~ m/s
and $\rho \ell$~= $3.75 \times 10^{-16} \Omega m^2$.
Eq.~\ref{jdp2} then yields for the Nb film $J_{dp,Nb}^{GL}(0) =
1.70 \times 10^{12}$~A/m$^2$, which is quite close to the
experimentally determined value of $J_{c0,Nb}^{GL}(0)) = 1.5
\times 10^{12}$~A/m$^2$ (see Table~\ref{param}). It is also in
good correspondence with the data presented by Ando {\it et al.}
\cite{ando93} on films with a thickness of 100~nm and different
strip widths between 0.1~$\mu m$ and 10~$\mu m$, who found a
fitted value $J_{dp,Nb}^{GL}(0) = 1.26 \times 10^{12}$~A/m$^2$. It
appears that the depairing current is directly probed by the
measurement of $J_{c0}$. \\
\noindent Next we consider the dependence of $J_{c0}^{GL}(0)$ on
the superconducting Nb layer thickness $d_{Nb}$. As Eq.~\ref{jdp2}
shows, $J_{dp}(0)$ is proportional to $\sqrt{ 1/\rho_{Nb} }$.
Since $\rho_{Nb}$ of the samples differs, this leads to some
variation in the expected value for $J_{dp}(0)$ which can be taken
into account by multiplying $J_{c0}^{GL}(0)$ by $\rho_{Nb}^{1/2}$.
Normalizing this value to the single Nb film yields the dependence
on $d_{Nb}$ as shown in Fig.~\ref{grafiek}. $J_{c0}^{GL}(0)$ in
the trilayers is clearly reduced with respect to the bulk Nb value
and increases with increasing $d_{Nb}$, but much more slowly than
$T_c$ does. The correspondence with theory is good at low
$d_{Nb}$, with some deviations above $d_{Nb} \approx 75$~nm. This
coincides with the findings on the temperature dependence of
$J_{c0}(t)$, shown in Fig.~\ref{JcJc0vT} : for small $d_{Nb}$
there are only small deviations in the whole measured temperature
regime, for large $d_{Nb}$ the deviations are large below $t =
0.9$. This suggests that at high $d_{Nb}$ the extrapolation for
$J_{c0}^{GL}(0)$ leads to somewhat underestimated values. In
essence, we conclude that the model used to describe the
depression of $T_c$ in F/S/S trilayers also adequately describes
the behaviour of $J_{dp}$. \\
This leaves one final point to be addressed. In order to determine
the depairing current it is usually understood that two conditions
have to be fulfilled \cite{romijn82,likharev79}~: the current has
to be distributed uniformly over the strip, and the width $w$
should be small enough to preclude vortex formation and motion. In
terms of penetration depth $\lambda(t)$, strip thickness $d_s$ and
Ginzburg-Landau coherence length $\xi(t)$ this means~:
\begin{eqnarray}
w < &\lambda_{eff}(t) &= \lambda(t) ,\;\;\;\;\;\;\;\;\;
d_s < \lambda_{eff}(t) \nonumber \\
    &             &= \lambda^2(t) / d_s , \;\; d_s > \lambda_{eff}(t)
\nonumber \\
w < &\,4.4 \xi(t) &= 4.4 \xi(0) \;/\; (1-t)^{1/2} \;.
\end{eqnarray}
Estimating $\lambda(0)$ from $\lambda(0)  = 1.05 \times 10^{-3}
\sqrt{\rho_0/T_c}$ we find it ranges between 67 and 113~nm. Both
conditions mean for all samples $1-t < 10^{-4}$, much smaller than
the region where $J_{c0}(t) \propto (1-t)^{3/2}$,
(Fig.~\ref{JcJc0vT}), and the question is valid whether the
current is uniform, as has implicitly been assumed in the
analysis. \\
Qualitatively, current is expected to pile up at the edges of the
strip in order to minimize the self-field inside. The edge current
will then sooner reach the value of $I_{dp}$. By using $J_{dp} =
I_{dp}/(wd)$, this would lead to underestimating the real value of
$J_{dp}$. From the close agreement between the experimental and
theoretical values this does not appear to be the case.
Quantitatively, the situation can be assessed that the self-field
of the sample is completely screened ($B_z =0$ in the sample). The
current distribution is then given by \cite{zeldov94}~:
\begin{equation}
J(x) = \frac{I_T}{\pi d \sqrt{W^2-x^2}} \;\;\;\;\;\;\; -W<x<W,
\label{eq:strip}
\end{equation}
where $I_T$ is the transport current through the sample, $x$ is in
the direction of the width $w$ of the film, $x=0$ in the middle of
the film and $2W=w$. According to this formula, the current
diverges at the edges of the film. It can be assumed, however,
that the field penetrates over a distance $d/2$ from the edges,
but is kept out of the rest of the sample by the screening
current. Then, the current within $d/2$ from the edges can be set
equal to $J_{dp}^{GL}(0)$ and beyond $d/2$ it decreases according
to Eq.~\ref{eq:strip}. The following calculation can be done for
the Nb film. The transport current $I_T$ in the screened part of
the strip can be calculated from Eq.~\ref{eq:strip} by using
\begin{equation}
J(x=W-d/2) =  J_{dp}^{GL}(0) \;.
\end{equation}
The total current $I$ including the edges is given by~:
\begin{eqnarray}
I/d &= &2 d/2 J_{dp}^{GL}(0) + \int_{-W+d/2}^{W-d/2}
       \frac{I_T}{\pi d \sqrt{W^2-x^2}} \nonumber \\
    &= &d J_{dp}^{GL}(0)\; \times \nonumber \\
    &  &(1 + \sqrt{\frac{w}{2d}} \: \left[ \sin^{-1} \frac{x}{|W|}
       \right]_{-W+d/2}^{W-d/2} )
\label{eq:current}
\end{eqnarray}
The ratio $I/(wd) \;/\; J_{dp}^{GL}(0)$, which can be calculated
from Eq.~\ref{eq:current}, gives the fraction of $J_{dp}^{GL}(0)$
which would be actually measured as the critical current under the
given current distribution, where the depairing current is reached
at the edges. It can be easily seen that it equals 1 when the
current is uniform. For the Nb film with $w$ and $d$ as given in
Table~\ref{param}, Eq.~\ref{eq:current} yields a fraction of 0.11,
an order of magnitude below what is actually measured. The
conclusion is that $J(x)$ is much more uniformly distributed than
might be expected. The reason is probably that a field and moving
vortices exist in the film, indicated by a voltage onset below the
jump to the normal state. This breaks up the Meissner state,
causes a much more uniform current distribution and allows the
correct determination of the depairing current over a much larger
region than expected on the basis of the condition $w < \lambda,
\xi$. Still, the deviations of $J_{c0}(t)$ compared to the
theoretical behaviour at higher $d_{Nb}$ in the Fe/Nb/Fe trilayers
may be due to the low value of $\lambda$ and inhomogeneities in
the current distribution at these high thicknesses. At low
thicknesses, there are two effects which increase $\lambda$ above
the bulk value. Firstly, for $d_s <$ $\lambda$ (around $d_{Nb}
\approx 75$~nm) the effective penetration depth increases
according to $\lambda_{eff} = \lambda(0)^2/d$, and can become
significantly higher than $\lambda(0)$. Secondly, the suppression
of the order parameter as measured by the decrease of $T_c /
T_{c0}$ results in a higher value for $\lambda(0)$. From that
point of view the full agreement between the measured and
calculated values of $J_{c0}(t)$ at the lowest thicknesses is not
surprising.

\section{Summary}
In this paper we have addressed the question of the value of the
superconducting depairing current in F/S/F trilayers with varying
$d_s$, where the superconducting order parameter is
inhomogeneously suppressed by the pair breaking in the F-layers.
The same model which is adequately describes the suppression of
$T_c$ with decreasing $d_s$ with two parameters (proximity
strength $\gamma$ and interface transparency $\gamma_b$ or $T$)
can also be used to compute the suppression of the depairing
current. Measurements of the zero-field critical current $J_{c0}$
(as defined by the current where the resistance jumps to the
normal state value) in thin strips of Fe/Nb/Fe show that the
temperature dependence near $T_c$ is as expected for the depairing
current. Also the absolute value of $J_{c0}$ of single Nb films is
close to the theoretically expected value and the measured
suppression of $J_{c0}$ in the trilayers follows the calculated
behaviour. We conclude that the current distribution is
homogeneous and that the depairing current is measured, even
though the strip widths are larger that the superconducting
penetration depth and coherence length. Also, the proximity effect
model correctly describes the shape of the order parameter, at
least in the superconducting layer. These findings can be of use
in experiments on the effect of injecting polarized
quasiparticles.

\section*{Acknowledgements}
This work is part of the research program of the 'Stichting voor
Fundamenteel Onderzoek der Materie (FOM)', which is financially
supported by NWO. A. A. G. acknowledges support by NWO in the
framework of the Russian-Dutch collaboration programme, grant
047-005-01. We would like to thank P. H. Kes for helpful
discussions.

\setcounter{equation}{0}
\renewcommand{\theequation}{A\arabic{equation}}
\section*{Appendix A : calculation of $J_{dp}$}  \label{appcalc}
We assume that the dirty limit conditions are fulfilled in both S
and F layers, so that the F/S bilayer can be described by the
Usadel equations. In the absence of a depairing current in the S
layer, and in the regime of large exchange energy in the
ferromagnet ($E_{ex}\gg k_BT_c$) these equations were discussed
extensively by Buzdin {\it et al.}\cite{buzdin92} (see also Demler
{\it et al.} \cite{demler97}). Here we rewrite these equation in $\theta $%
-parametrization ($F=\sin \theta $, $G=\cos \theta $) and include
the pair-breaking effects by current along the S film~:

\begin{equation}
\xi _S^2\frac{d^2}{dz^2}\theta _S(z)-\widetilde{\omega }\sin
\theta _S(z)+\Delta (z)\cos \theta _S(z)=0,
\end{equation}

\begin{equation}
\xi _F^2\frac{d^2}{dz^2}\theta _F(z)-i\sin \theta _F(z)=0,
\end{equation}

\begin{equation}
\Delta \ln (T/T_c)+\pi T\sum_{\omega _n}(\frac \Delta {\left|
\omega _n\right| }-\sin \theta _S)=0,
\end{equation}
where $\omega _n=\pi (2n+1)T/T_c$ is the normalized Matsubara frequency,  $%
\widetilde{\omega }=\left| \omega _n\right| +Q^2\cos \theta (z)$,
$\Delta $ is the pair potential in a superconductor normalized to
$\pi T_c$, $\xi_S=(\hbar D_S/2\pi T_c)^{1/2}$, $\xi _F=(\hbar
D_F/2E_{ex})^{1/2}$ and $D_{F,S}$ are the coherence lengths and
the electronic diffusion coefficients in F and S metals. Moreover,
$Q=\xi _s\partial \chi /\partial x$ is the normalized
gradient-invariant superfluid velocity in the $x$-direction, with
$\chi $ the phase of the pair potential $\Delta $. There are two
sources of pair-breaking in the problem, the volume one by the
current and the surface one by the ferromagnet. The latter is
described by the boundary conditions at the FS interface ($z=0$)
\begin{eqnarray} \label{eq:bound}
\xi_S\frac d{dz}\theta _S & = & \gamma \xi _F\frac d{dz}\theta _F
\\
\gamma _b\xi _F\frac d{dz}\theta _F & = & \sin (\theta _S-\theta
_F)
\end{eqnarray}
where the parameter $\gamma =\rho _S\xi _S/\rho _F\xi _F$
describes the strength of the suppression of superconductivity in
S by the ferromagnet.

The parameter $\gamma _b$ describes the effect of boundary
transparency (coupling strength) between the layers. In the NS
case, when the decoupling is due to the presence of an additional
potential barrier at the interface, $\gamma _b=R_B/\rho _F\xi _F$,
with $R_B$ the normal state resistance of the N/S interface
\cite{kupr88}. In the F/S bilayer there is no general microscopic
derivation for $\gamma _b$, combining the effect of exchange
splitting and an additional interface barrier. A simple estimate
is still possible, when the exchange splitting is the main cause
for intransparency \cite{aarts97}. Then $\gamma _b=(2/3)(l_F/\xi
_F)<(1-T_A)/T_A>$ where $T_A$ is the transmission probability of
scattering between the majority and minority spin subbands, i.e.
the probability of Andreev reflection. This process is implicitly
described by the boundary condition $\gamma _b\xi _F\frac
d{dz}\theta _F=\sin (\theta _S-\theta _F)$ since $\theta _F$ is
off-diagonal in spin indices. Here the brackets $<...>$ denote the
Fermi surface averaging, which is generally proportional to the
overlap area of the projections of different spin subbands onto
the contact plane \cite{dejong95,mazin99}. As a result, $T_A$
drops roughly linearly (for spherical Fermi surfaces) as a
function of $E_{ex}$, both for ballistic and diffusive interfaces
\cite{mazin00}. The supercurrent density is given by
\begin{equation}
J_s(z,Q)=\frac{2\pi \sigma _s}eQT\sum_{\omega _n}\sin ^2\theta _s
\;. \label{A6}
\end{equation}
Since the superconducting pair potential $\Delta $ and the Green's function $%
\theta _s$ are suppressed by the superflow $Q$, the dependence
$J_s(Q)$ must be found selfconsistently. In the well known
spatially homogeneous case \cite{Bardeen62} the function $J_s(Q)$
behaves non-monotonously : the supercurrent $J_s$ increases with
$Q$ at small $Q$, then reaches a maximum and finally drops to
zero, when $\Delta $ is fully suppressed by current. The depairing
current is defined as the maximum of $J_s(Q)$. A similar situation
holds in the spatially inhomogeneous case considered here, with
the difference that the solutions for $\theta(z)$ and $\Delta(z)$
of the proximity effect problem (eqs.~A1-A3)  should be calculated
selfconsistently for a given $Q$ using the boundary conditions at
the FS interface (eqs.~A4, A5). This problem is solved numerically
by the method applied previously to NS bilayers and described in
detail in \cite{Gol95}. Then the local $z$-dependent supercurrent
density $J_s(z,Q)$ is calculated from eq.A6 by summing the
solutions $\sin ^2\theta _s$ over $\omega_n$. Finally the density
is averaged over film thickness $J_s(Q) =$
$d_s^{-1}\int_0^{d_s}J_s(z,Q) dz$ and the depairing current is
found from the maximum of the dependence of $\left\langle
J_s\right\rangle $ on $Q$.

\end{document}